\documentclass[prl,twocolumn,showpacs,superscriptaddress,nofootinbib]{revtex4-2}
\pdfoutput=1
\usepackage{graphicx}
\usepackage{epstopdf}
\usepackage{bm}
\usepackage{amssymb}
\usepackage{color}
\usepackage{amsmath}
\usepackage{amstext}
\usepackage{latexsym}
\usepackage[usenames,dvipsnames]{xcolor}
\usepackage[colorlinks=true,citecolor=MidnightBlue,linkcolor=RubineRed,urlcolor=MidnightBlue]{hyperref}
\usepackage{float}

\def\x{{\rm\bf x}}

\newcommand{\beq}{\begin{equation}}
\newcommand{\eeq}{\end{equation}}
\newcommand{\beqa}{\begin{eqnarray}}
\newcommand{\eeqa}{\end{eqnarray}}

\usepackage{tikz,xcolor,hyperref}
\definecolor{lime}{HTML}{A6CE39}
\DeclareRobustCommand{\orcidicon}{
\begin{tikzpicture}
\draw[lime, fill=lime] (0,0)
circle[radius=0.16]
node[white]{{\fontfamily{qag}\selectfont \tiny \.{I}D}}; 
\end{tikzpicture}
\hspace{-2mm}
}
\foreach \x in {A, ..., Z}{%
\expandafter\xdef\csname orcid\x\endcsname{\noexpand\href{https://orcid.org/\csname orcidauthor\x\endcsname}{\noexpand\orcidicon}}
}

\begin{document}

\title{Macroscopic Efimov effect of quantized vortex}

\author{Wei-Can Yang\hspace{-1.5mm}\orcidA{}
}\email{weicanyang@outlook.com}
\affiliation{Department of Physics, Osaka Metropolitan University, 3-3-138 Sugimoto, 558-8585 Osaka, Japan}
\affiliation{Center for Theoretical Physics , Hainan University, Haikou 570228, China}

\author{Makoto Tsubota}
\affiliation{Department of Physics, Osaka Metropolitan University, 3-3-138 Sugimoto, 558-8585 Osaka, Japan}
\affiliation{Nambu Yoichiro Institute of Theoretical and Experimental Physics (NITEP), Osaka Metropolitan University, 3-3-138 Sugimoto, Sumiyoshi-ku, Osaka 558-8585, Japan}

\author{Muneto Nitta\hspace{-1.5mm}\orcidC{}}
\affiliation{Research and Education Center for Natural Sciences, Keio University,
Hiyoshi 4-1-1, Yokohama, Kanagawa 223-8521, Japan}
\affiliation{
International Institute for Sustainability with Knotted Chiral Meta Matter(SKCM$^2$), Hiroshima University, 1-3-2 Kagamiyama, Higashi-Hiroshima, Hiroshima 739-8511, Japan
}

\author{Hua-Bi Zeng\hspace{-1.5mm}\orcidB{}
}\email{zenghuabi@hainanu.edu.cn}
\affiliation{Center for Theoretical Physics , Hainan University, Haikou 570228, China}

\begin{abstract}
The three-body problem, from the chaotic motions of celestial bodies to complex microscopic particle interactions, has always been one of the most foundational yet intricate challenges in physics since its establishment.  A key breakthrough in this domain is the Efimov effect, which represents a significant stride in what is now known as Efimov physics. Our study uncovers a macroscopic Efimov effect in a three-component Bose-Einstein Condensate (BEC) system.  Through theoretical analysis and numerical simulation, it is verified that under certain conditions, three vortices form a bound state, while removing one vortex causes the others to unbind, demonstrating topological characteristics similar to the Borromean rings, hence termed the `vortex Efimov effect', signifying a novel topological phase transition.  We propose several experimental approaches to realize this macroscopic Efimov effect, paving new paths not only in many-body physics but also in exploring quantum phase transitions and applications in quantum information.
\end{abstract}

\maketitle
\begin{figure}[t]
    \centering
   \includegraphics[width=8.5cm,trim=50 20 0 20]{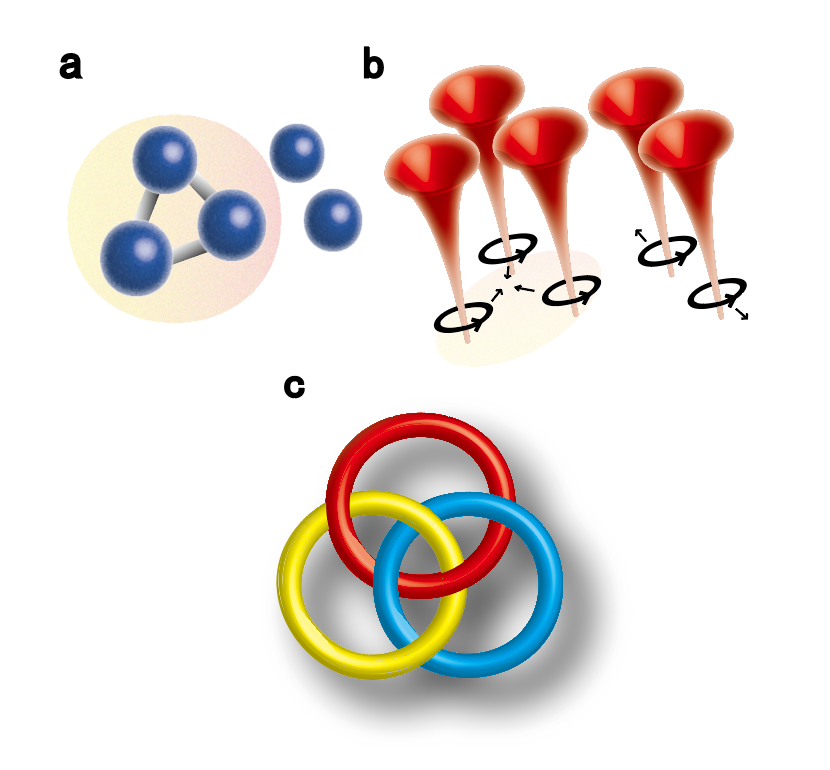}
    \caption{Illustration of the vortex Efimov effect:
{\bf a}: The Efimov effect in microscopic particles: three particles form a bound state, while any two particles remain unbound.
{\bf b}: The corresponding vortex Efimov effect: three vortices attract each other, while any two vortices repel each other.
{\bf c}: The topological structure shared by both effects—Borromean rings: each ring is intertwined with the other two, and if one ring is removed, the remaining two are no longer connected.}
    \label{figure1}
\end{figure}

{\it Introduction}
The study of the three-body problem dates back to the establishment of the cornerstones of physics - Newtonian mechanics. To this day, it remains a significant challenge in physics, so much so that most research is constrained to the simplifying assumption of two-body interactions \cite{Burnett_2002,Kohler2006,Pethick_2008}. This limitation highlights the unique position of the three-body problem in theoretical physics and the difficulty of addressing such complex issues.
Until 1970s, when Efimov proposed the concept of effective three-body interactions among identical particles and predicted the weakly bound states of three neutral bosons, even in the absence of two-body binding (Fig.\ref{figure1} a) - now known as the Efimov effect \cite{Efimov1970,Efimov1970_2}, physics entered the era of a universal theory for three-body interactions among microscopic particles.

More than three decades later, this effect was experimentally discovered in ultracold atoms, reinvigorating interests and leading to a surge of research by experimental physicists \cite{Kraemer_2006,Knoop_2009,Zaccanti_2009,Pollack_2009,Bo2014,Tung2014,Johansen_2017,Pires2014,Kunitski_2015}. This renewed interest gradually led to the development of a distinct and universal Efimov physics \cite{Naidon_2017}. However, despite the active research on the Efimov effect within the platform of condensed matter systems, its impact has been primarily confined to the development of 
atomic physics and has not significantly influenced phenomena within condensed matter physics itself. Nevertheless, the universality of the Efimov effect suggests that it should also manifest in condensed matter systems, which would require the presence of effective three-body interaction potentials described by \cite{B_chler_2007}:
\begin{align}
    V_{eff}=\sum_{i<j}V({\bf r}_i-{\bf r}_j)+\sum_{i<j<k}U({\bf r}_i,{\bf r}_j,{\bf r}_k)
\end{align}
Therefore, an ideal system for such studies is the Bose-Einstein condensate (BEC) of cold atoms with multi-body interactions. This choice is not only because the Efimov effect is realized within such systems, but also because they naturally incorporate 
the stable and well-defined macroscopic quantum elements - quantized vortices \cite{Esry1999,Fetter2009}.

Quantized vortices possess both microscopic and macroscopic properties. In terms of formation, they can only appear in quantum phases that break gauge symmetry. From the perspective of dynamics and observation, they exhibit some attributes typical of macroscopic objects.  This dual nature allows quantized vortices to serve as a unique bridge between the quantum and classical worlds. 
Therefore, we expect that quantized vortices can bridge microscopic three-body Efimov interactions with macroscopic phenomena. To distinguish individual vortices, we consider a coupled system of three identical components BEC, focusing on the interactions among three vortices under three-body interaction. 
In this artile, we discover that, within a tunable range of interactions, the trio of vortices exhibits a topological structure analogous to Efimov states as illustrated in Fig.\ref{figure1} b. When all three vortices are present, a tripartite bound state forms. However, removing one vortex results in the remaining two becoming unbound, which aligns with the topological structure of Borromean rings (Fig.\ref{figure1} c).

To demonstrate the three-body interaction between three-component BECs, we consider the energy with three macroscopic wave function \cite{Bedaque2009}:
\begin{equation}
\begin{aligned}
E(\Psi)&=\int d^3x [ K(\Psi)+V(\Psi)]\\
K(\Psi)&= \sum_{i=1,2,3}(-\frac{\hbar^2}{2m_i}\Psi_i^*\nabla^2\Psi_i)\\
V(\Psi)&= \sum_{i=1,2,3}\frac{g_i}{2}|\Psi_i|^4+ \sum_{i,j=1,2,3}\frac{g_{ij}}{2}|\Psi_i|^2|\Psi_j|^2\\&+g_{123}|\Psi_1|^2|\Psi_2|^2|\Psi_3|^2.
\end{aligned}
\end{equation}

Based on general considerations, we assume the three components are different hyperfine states of the same atom, so the atomic masses $m_i$ and the interatomic interactions $g_i$  are all equal and represented as $m$ and $g$. Here $g_{ij}$ and $g_{123}$  represent the two-body and three-body interaction coefficients, respectively. To simplify the model, we have focused only on the three-body interactions between the three components, because intra-component three-body effects do not significantly influence the outcomes, and experimental control over the coefficients is only feasible between different components.

By applying the variational principle $i\hbar\partial_t\Psi_i=\delta E/\delta \Psi_i^*$, we can derive the equations of motion from the energy functional, resulting in the coupled three-component Gross-Pitaevskii equations that include three-body interactions
\begin{align}
     &i\hbar\partial_t\Psi_i= \nonumber\\
     &\big( -\frac{\hbar^2\nabla^2}{2m} +g|\Psi_i|^2+\eta(|\Psi_j|^2+|\Psi_k|^2)+\epsilon|\Psi_j|^2|\Psi_k|^2 \big)\Psi_i \nonumber\\
     &(i, j, k) \text{ is a cyclic permutation of } (1, 2, 3)
\label{EoM}
\end{align}
where, to ensure the identical nature of the interactions among the three vortices, we choose $g_{ij}=\eta$ and $g_{123}=\epsilon$.

Further by consider a time-dependence $\Psi_i({\bf x},t)=e^{-i\mu t/\hbar}\Psi_i({\bf x })$ with same chemical potential in three component, we can obtain the stationary equation with the ground state wave function
\begin{align}
    &\mu \Psi_i= \nonumber \\
     &\big( -\frac{\hbar^2\nabla^2}{2m} +g|\Psi_i|^2+\eta(|\Psi_j|^2+|\Psi_k|^2)+\epsilon|\Psi_j|^2|\Psi_k|^2 \big)\Psi_i,  \nonumber\\
    &|\Psi_i|^2 \equiv v^2 =\frac{\sqrt{4\epsilon\mu+(2\eta+g)^2}-(2\eta+g)}{2\epsilon}.
\label{stationary}
\end{align}

The topology of this ground state is characterized $\pi_1[U(1)^3]=\mathbb{Z}\oplus\mathbb{Z}\oplus\mathbb{Z}$. It allows three kinds of winding numbers, refer to $(1,0,0)$, $(0,1,0)$, $(0,0,1)$. When
moving around the $(1,0,0)[(0,1,0),(0,0,1)]$ vortex, the phase of $\Psi_1(\Psi_2,\Psi_3)$
rotates by $2\pi$, while the phases of the other components remain unchanged. Similarly, we can obtain representations of multi-vortex structures such as $(1,1,0)$ and $(1,1,1)$.
A set of these three vortices $(1,1,1)$ carries a unit circulation while each carries a 1/3 quantized circulation \cite{Eto:2012rc,Eto:2013spa}.

First, as an ansatz for the vortex structure, we can write the axisymmetric form of a single vortex $(1,0,0)$ in different components:
\begin{equation}
    \begin{aligned}
    \Psi_1^{(1,0,0)}&=ve^{i\theta}f_{(1,0,0)}(r)\\\
    \Psi_2^{(1,0,0)}&=vh_{(1,0,0)}(r)\\
    \Psi_3^{(1,0,0)}&=vl_{(1,0,0)}(r) 
\end{aligned}\label{profile}
\end{equation}
where $r$ and $\theta$ are the polar coordinates. The profile functions $f_{(1,0,0)}$, $h_{(1,0,0)}$ and $l_{(1,0,0)}$ can be obtained by substituting vortex configuration function in 
Eq.~(\ref{profile}) into stationary Eq.~(\ref{stationary}) with the boundary condition $(f_{(1,0,0)},h_{(1,0,0)},l_{(1,0,0)})\rightarrow (1,1,1)$ as $r \rightarrow \infty$ and $(f_{(1,0,0)},h'_{(1,0,0)},l'_{(1,0,0)})\rightarrow (0,0,0)$ 
as $r \rightarrow 0$, with the prime denoting a differentiation with respect to $r$.

We adopt a common and practical vortex structure as a trial solution and improved the profile function trial solutions for other components based on this, satisfying the boundary conditions with the fullest degrees of freedom \cite{Pethick_2008}:
\begin{equation}
\begin{aligned}
    f_{(1,0,0)}(r)&=\frac{r}{\sqrt{a+r^2}}\\
    h_{(1,0,0)}(r)&=1+c-\frac{cr}{\sqrt{b+r^2}}\\
    l_{(1,0,0)}(r)&=1+c-\frac{cr}{\sqrt{b+r^2}}.
\end{aligned}
\end{equation}
\begin{figure*}[t]
    \centering
   \includegraphics[width=16cm,trim=40 50 0 0]{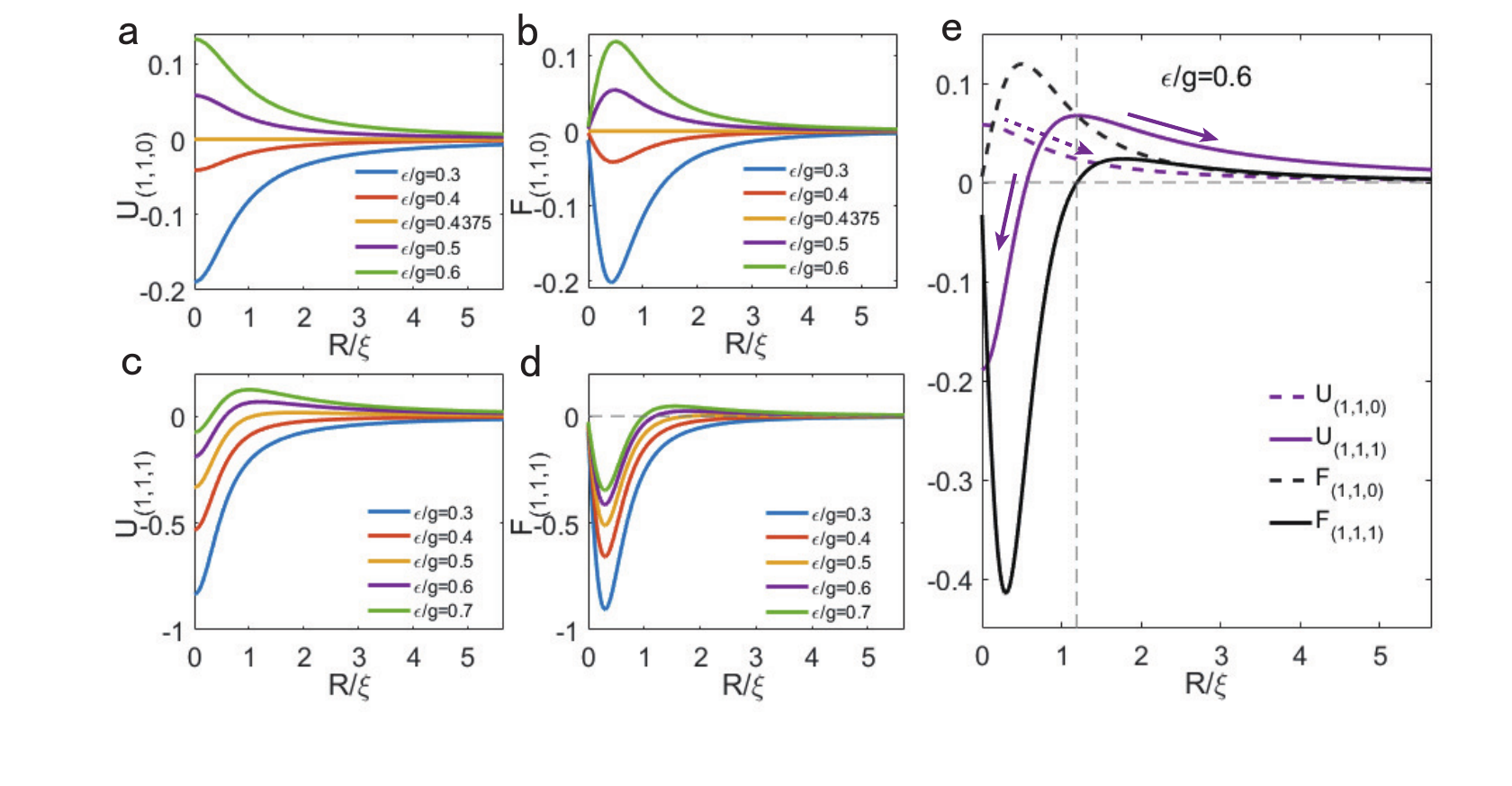}
    \caption{Vortex interaction potential and force as the function of distance. {\bf a,b}: The interaction  between two vortices. From top to bottom, the three-body interaction decreases progressively, and when $p^-=0$, that is $\epsilon=0.4375g$, no interaction between two vortices. {\bf c,d}: 
For the same parameters, the interaction potential of the three vortices is no longer monotonic when $p^- > 0$. When the vortex distance is less than the critical value, the three vortices attract each other; when it is greater than a critical value, they repel each other. This corresponds to the interaction force below and above the dashed line in d.
{\bf e}: An interaction diagram displaying the vortex Efimov effect when $\epsilon=0.6g$. The arrows indicate the direction of state motion in the potential field. In the two-vortex case, the arrows always point to the right, indicating repulsion between the vortices. In the three-vortex case, to the right of the critical distance marked by the vertical dashed line, the arrows point to the right, indicating repulsion; to the left of the critical distance, the arrows point to the left, towards a vortex coinciding state, indicating attraction between the vortices. This represents the vortex Efimov state, where one vortex is intertwined with the other two vortices, forming a Borromean ring structure.}
\label{figure2}
\end{figure*}
In this way, we obtain the correct vortex structure function in the three-component BEC,
\begin{equation}
\begin{aligned}
 a&=\frac{p^+}{(p^+-2p^-)(p^++p^-)}\frac{\hbar^2}{2v^2m}\\
b\times c&=\frac{p^-}{(p^+-2p^-)(p^++p^-)}\frac{\hbar^2}{2v^2m}
\end{aligned}
\end{equation}
where $p^+=g+\eta+v^2\epsilon$ and $p^-=\eta+v^2\epsilon$.
From this structure, we can obviously find that the instability condition shifts from the two-component phase separation $g < \eta$ \cite{Kasamatsu2003} to the inclusion of three-body interactions in $g<\eta+v^2\epsilon$. 
If $g>\eta+v^2\epsilon$, $a$ is always positive while sign of 
$b\times c$ depends on parameter $p^-=v^2\epsilon+\eta$.  From profile function $h_{(1,0,0)}$ and $l_{(1,0,0)}$, $c$ should be a small value and $b$ should be a positive value.
Since $v^2=\rho_s$ is the background density, so we know the profile function of unwinding component at the vortex center is concave for $\rho_s\epsilon+\eta < 0$ and convex for the $\rho_s\epsilon+\eta > 0$.
Profile functions of vortices in the other components $\Psi^{(0,1,0)}$ and $\Psi^{(0,0,1)}$ can be obtained by using same process.

With these structure functions, we can proceed to calculate the actual interactions of two and three vortices. Considering the vortex placements symmetric with respect to the coordinate axis center, the two vortices $(1,0,0)$ and $(0,1,0)$ are located at $(x,y)=(R,0)$ and $(x,y)=(-R,0)$, respectively. The third vortex is added at the position $(x,y)=(0,\sqrt{3}R)$, forming a $C_3$ rotationally symmetric structure in the case of three vortices.

In two vortex case, the interaction potential can be obtained by subtracting two individual energies from the total energy as
\begin{align}
    U_{(1,1,0)}=\int d^2x(\delta K_{(1,1,0)}+\delta V_{(1,1,0)})
\end{align}
where two contributions form kinetics energy $\delta K_{(1,1,0)}=K(\Psi_i^{(1,1,0)})-K(\Psi_i^{(1,0,0)})-K(\Psi_i^{(0,1,0)})$ and interaction potential energy $\delta V_{(1,1,0)}=V(\Psi_i^{(1,1,0)})-V(\Psi_i^{(1,0,0)})-V(\Psi_i^{(0,1,0)})$, where $\Psi_i^{(1,1,0)}$ come from the standard Abrikosov ansatz $\Psi_1^{(1,1,0)}=v^{-1}\Psi_i^{(1,0,0)}\Psi_i^{(0,1,0)}$ \cite{Eto2011}.
Then, by using the derivation of the interaction potential with respect to the distance, one can then calculate the interaction force
\begin{align}
     F_{(1,1,0)}=-\frac{\partial U_{(1,1,0)}}{2\partial R}.
\end{align}

For the three-vortex case, we repeat the above steps, but consider replacing one of the vortices with a vortex pair,
\begin{equation}
    \begin{aligned}
U_{(1,1,1)}&=\int d^2x(\delta K_{(1,1,1)}+\delta V_{(1,1,1)})\\
F_{(1,1,1)}&=-\frac{\partial U_{(1,1,1)}}{3\partial R}
\end{aligned}
\end{equation}
with contributions form kinetics energy $\delta K_{(1,1,1)}=K(\Psi_i^{(1,1,1)})-K(\Psi_i^{(1,1,0)})-K(\Psi_i^{(0,0,1)})$ and interaction potential energy $\delta V_{(1,1,1)}=V(\Psi_i^{(1,1,1)})-V(\Psi_i^{(1,1,0)})-V(\Psi_i^{(0,0,1)})$, where $\Psi_i^{(1,1,1)}=v^{-2}\psi_i^{(1,0,0)}\Psi_i^{(0,1,0)}\Psi_i^{(0,0,1)}$.

We start with a simple assumption: to facilitate calculations, we assume the small quantity 
$c$ is $0$, so that the vortex does not cause fluctuations in other components with $h=l=1$. 
In this way, we can relatively easily calculate the interactions,
\begin{equation}
    \begin{aligned}
        U_{(1,1,0)}&=a^2 \pi v^4(\eta+v^2\epsilon)\frac{ \mathrm{Arctanh}(R/\sqrt{R^2+a})}{R\sqrt{R^2+a}}\\
        F_{(1,1,0)}&=-1/2a^2 \pi v^4(\eta+v^2\epsilon)\times\\ &\frac{R\sqrt{a+R^2}-(a+2R^2)\mathrm{Arctanh}(\frac{R}{\sqrt{a+R^2}})}{R^2(a+R^2)^{3/2}}
        \end{aligned}
\end{equation}
We observe that in the case of two vortices, the vortex interaction depends only on $p^-=\eta+v^2\epsilon$. 
To maintain generality, we use dimensionless parameters $g= 7$, $\mu = 4$, and fix $\eta=-0.5g$ with the scales of length and time by $\xi=\sqrt{\hbar^2/2m\mu}$ and $\tau=\hbar/\mu$ \cite{Kasamatsu2016}.
As shown in Fig. \ref{figure2} (a and b),  as $\epsilon$ gradually increases, the potential energy keeps negative when $p^- < 0$, representing an attractive force between the two vortices. When $p- > 0$, that is $\epsilon=0.4375g$, the potential energy becomes positive, and the two vortices repel each other.
This indicates that the interaction direction between the two vortices does not change with distance and is monotonic.

However, for three vortex case, we find that the potential energy is 
\begin{equation}
    \begin{aligned}
        U_{(1,1,1)}&=\int d^2x v^4[\eta(1-k_3)(1-k_1)+\eta(1-k_3)(1-k_2)\\
        &+v^2\epsilon(1-k_1k_2)(1-k_3)]
    \end{aligned}
\end{equation}
where $k_{(1,2,3)}=\frac{r_{(1,2,3)}^2}{a+r_{(1,2,3)}^2}$ and $r_{i}$ refers to the coordinate position relative to vortex $i$.
This integral is difficult to solve analytically, but from its form, we can see that it is no longer a monotonic function dependent only on $p^-$.  We can use high-precision discrete methods for numerical integration to simultaneously obtain the interaction force.
\begin{figure}[t]
    \centering
   \includegraphics[width=7cm,trim=0 0 0 0]{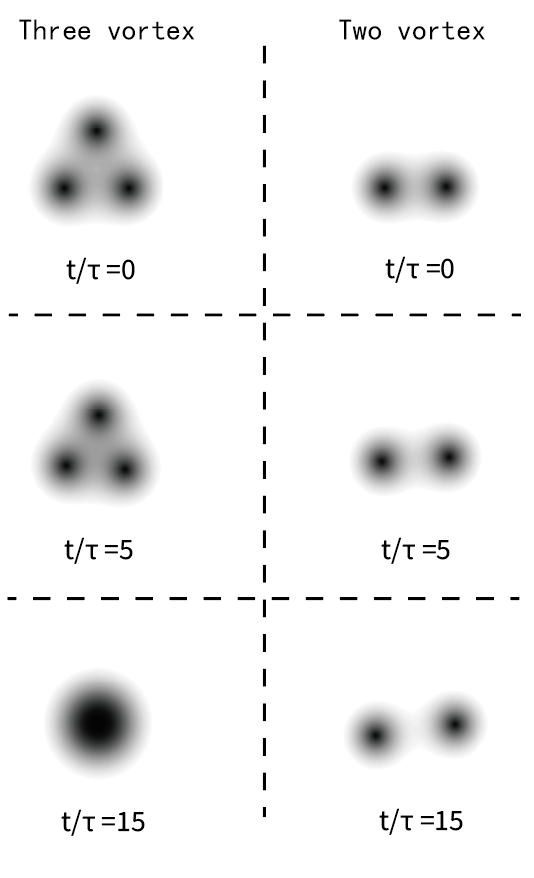}
    \caption{The dynamic evolution of vortex Efimov states with $\epsilon = 0.6g$ and $R/\xi = 1$. On the left, the evolution of three vortices, initially positioned symmetrically with equal relative distances. On the right, the evolution of two vortices, also initially positioned symmetrically with equal relative distances. Ultimately, the three vortices approach each other to form a giant vortex, while the two vortices completely separate.}
    \label{figure3}
\end{figure}
\begin{figure}[t]
    \centering
   \includegraphics[width=12cm,trim=30 30 0 20]{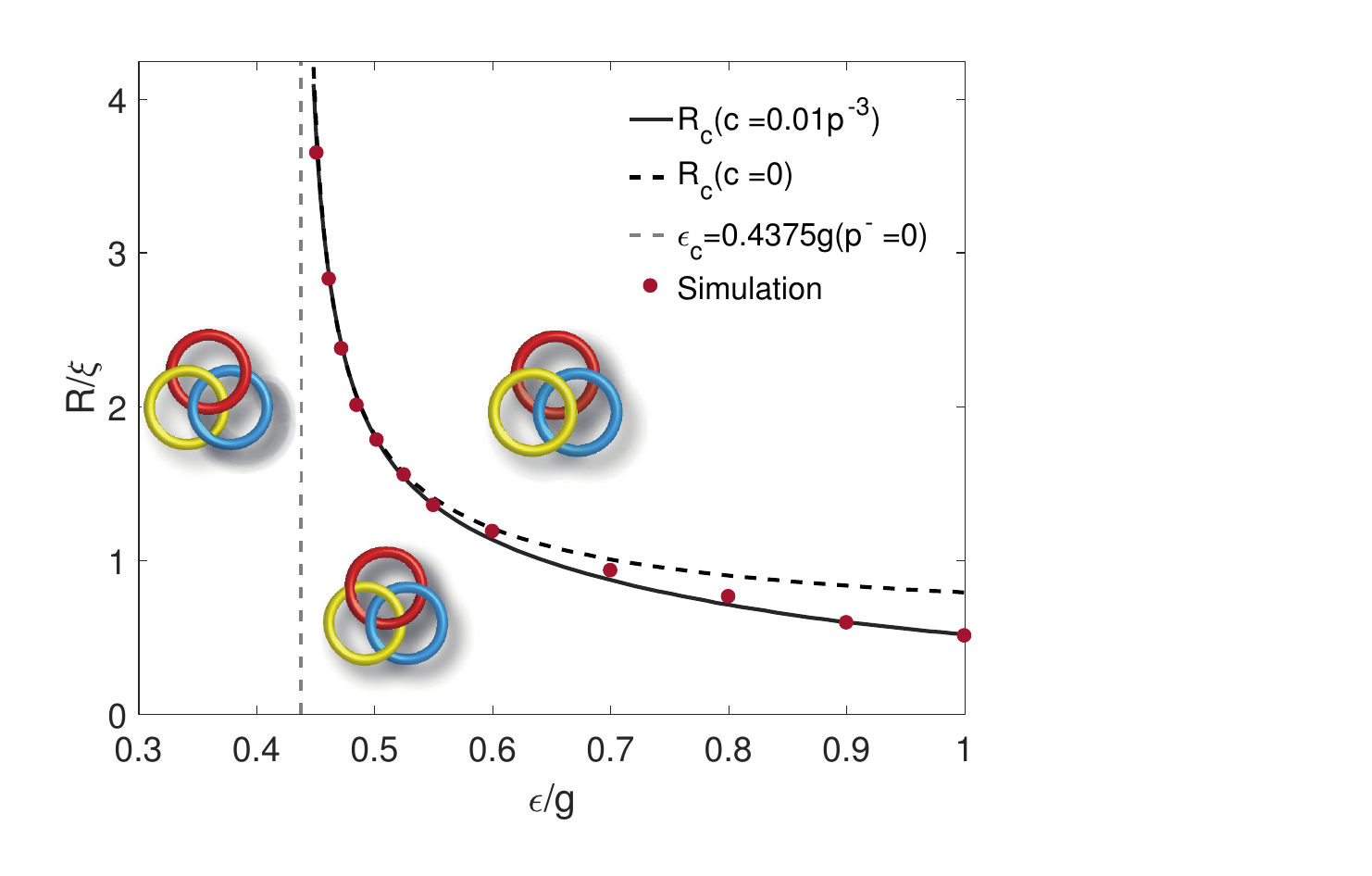}
    \caption{Phase diagram of vortex Efimov effect. The structure on the left is the fully bound three-ring structure, the bottom shows the Borromean rings(vortex Efimov state), and the right depicts the completely unbound three rings. The dashed curve represents the ideal vortex calculation results with $c=0$, the solid curve represents the actual vortex calculation results with the practical assumption $c=0.01 p^{-3}$, and the red circles indicate the numerical simulation results.}
    \label{figure4}
\end{figure}

Fig.~\ref{figure2} (c,d) shows the interaction potential energy and interaction force for three vortices, which completely corresponds to the parameter for two-vortex case in Fig.~\ref{figure2} (a,b).
It can be observed that in the case of three vortices, when $p^- > 0$, there is a critical distance $R_c$. Below this critical distance, the vortices attract each other; above this critical distance, the vortices repel each other. In Fig.\ref{figure2} e, we selected one of the parameters $\epsilon=0.6g$ to compare the interactions between two vortices and three vortices. 
On the left side of the critical distance indicated by the dashed line, the two vortices repel each other while the three vortices attract each other. This indicates the emergence of a new vortex state with a Borromean topological structure consistent with the Efimov effect, which we term it as the ``vortex Efimov effect.''

By performing numerical simulations of the dynamic evolution of the equations of motion in Eq.~(\ref{EoM}), the theoretical results can be validated. In Fig.~\ref{figure3} and movie~\cite{Movie}, we present the actual simulation of vortex dynamics with parameter 
$\epsilon=0.6g$ and initial distance $R/\xi=1$. It can be observed that for the same initial distance, the three vortices gradually attract each other and form a multi-winding giant vortex, while the two vortices gradually move apart until there is no interaction.

By repeatedly performing numerical simulations with varying initial vortex distances, we can obtain the phase diagram of the vortex Efimov state, showing the variation of the critical distance with three-body interactions, as shown in Fig.~\ref{figure4}. On the left side of the critical $p^-_c$ ($\epsilon_c=0.4375g$), there is the fully bound state phase where the three vortices (rings) are completely intertwined, and even if one vortex (ring) is removed, the remaining two are still entangled. 
On the right side of the critical $p^-_c$, there are two phases: when the distance is greater than the critical value, it is the completely unbound state phase; when the distance is less than the critical value, it is the vortex Efimov state, where the binding of each pair of vortices (rings) is mediated by the third vortex (ring). If one vortex (ring) is removed, the remaining two are unbound.

From the critical distance we can find that when $p^-$ is small, the theoretical values under the $(c = 0$ approximation match the actual simulation values well. However, when $p-$ is too large, this approximation is no longer valid, and therefore we need to consider a finite $c$. A reasonable 
ansatz can be made here, assuming $c$ is a small value and proportional to $p^-$, then $c$ and $b$ should become
\begin{equation}
    \begin{aligned}
        c&=tp^{-(d)}\\
    b&=\frac{1}{t}\frac{1}{p^{-(d-1)} (p^+-2p^-)(p^++p^-)}\frac{\hbar^2}{2v^2m}
    \end{aligned}
\end{equation}
when we set $t = 0.01$ and $d = 3$, the theoretical values perfectly match the numerical simulation results. Consequently, we obtained a more accurate vortex structure and a complete phase diagram of the topological phase transition of the vortex Efimov state.

{\it Discussion.}
Our study has successfully demonstrated the vortex Efimov effect in a three-component BEC system through both theoretical analysis and numerical simulation. One promising approach to experimentally realize the vortex Efimov state is to utilize a polar molecules system where natural strong three-body interactions exist \cite{B_chler_2007}. By fine-tuning the Feshbach resonances between each pair of components, one can control the interaction strengths and achieve the conditions necessary for the formation of vortex Efimov states \cite{Inouye_1998,Courteille1998}. 
Another potential candidate is cold atom lithium-6, where experiments have observed three-component Efimov effects \cite{Nakajima2010,Nakajima2011,Williams2009}. This makes it an ideal candidate for studying three-component vortex Efimov effects. By tuning the three-body and two-body parameters with magnetic field, vortex Efimov states may be achieved.


Future research can delve into the unique properties of three-dimensional vortex lines in vortex Efimov states or the Efimov state conditions of more then three components, potentially uncovering distinctive topological features and dynamic behaviors. Moreover, studying the dynamics of these vortex states offers a new framework for investigating many-body Efimov physics, enhancing our understanding of complex interactions in quantum systems.
The unique topological characteristics of vortex Efimov states, such as the Borromean ring structure, may have significant applications in quantum information processing. These states could be utilized to develop robust quantum memory systems and advanced quantum communication protocols, leveraging their topological stability and resilience to perturbations.

\section{Method}

\subsection{Vortex configuration}
The initial vortex distribution can be prepared using a polar coordinate system. For example, consider two well-separated vortex $(1,0,0)$ and $(0,1,0)$ in different components.

Lets place the $(1,0,0)$ and $(0,1,0)$ vortices at $(x,y)=(R,0)$ and $(x,y)=(-R,0)$, respectively. We
use the polar coordinates $(r,\theta)$ with the origin $(x,y) = (0,0)$ and express the relative coordinates from $(1,0,0)$ and $(0,1,0)$ vortex center as $[r_{(1,0,0)},\theta_{(1,0,0)}]$ and $[r_{(0,1,0)},\theta_{(0,1,0)}]$.
Then we have
\begin{equation}
\begin{aligned}
    &r_i^2=(r\ \mathrm{cos}\theta\mp R)^2+r^2\mathrm{sin}^2\theta=(x\mp R)^2+y^2\\
    &\mathrm{tan}\theta_i=\frac{r\ \mathrm{sin}\theta}{r\ \mathrm{cos}\theta\mp R}=\frac{y}{x\mp R}
\end{aligned}
\end{equation}
with $i = 1 \ or \ (1,0,0),\  2 \  or \ (0,1,0)$, the minus sign for $i = 1 \ or \ (1,0,0)$, and the
plus sign for $i =2 \ or \ (0,1,0)$.
Then the vortex configurations can be expressed as
\begin{align*}
    \psi_1^{(1,0,0)}&=ve^{i\theta_{(1,0,0)}}f_{(1,0,0)}[r_{(1,0,0)}]=ve^{i\theta_{1}}\left(\frac{r_1}{\sqrt{a+r_1^2}}\right)\\
 \psi_2^{(1,0,0)}&=vh_{(1,0,0)}[r_{(1,0,0)}]=v\left(1+c-\frac{cr_1}{\sqrt{b+r_1^2}}\right)\\
 \psi_3^{(1,0,0)}&=vl_{(1,0,0)}[r_{(1,0,0)}]=v\left(1+c-\frac{cr_1}{\sqrt{b+r_1^2}}\right)\\
  \psi_1^{(0,1,0)}&=vh_{(0,1,0)}[r_{(0,1,0)}]=v\left(1+c-\frac{cr_2}{\sqrt{b+r_2^2}}\right)\\
   \psi_2^{(0,1,0)}&=ve^{i\theta_{(0,1,0)}}f_{(0,1,0)}[r_{(0,1,0)}]=ve^{i\theta_{2}}\left(\frac{r_2}{\sqrt{a+r_2^2}}\right)\\
   \psi_3^{(0,1,0)}&=vl_{(0,1,0)}[r_{(0,1,0)}]=v\left(1+c-\frac{cr_2}{\sqrt{b+r_2^2}}\right).
\end{align*}

Similarly, the third vortex $(0,0,1)$ can be placed at $(x, y) = (0, \sqrt{3})$. Thus, the three vortices form a $C_3$ rotationally symmetric structure,
\begin{equation}
\begin{aligned}
    &r_{(0,0,1)}^2=r^2\mathrm{sin}^2\theta+(r\mathrm{cos}\theta-\sqrt{3}R)^2\\
&\mathrm{tan}\theta_{(0,0,1)}=\frac{r\mathrm{cos}\theta-\sqrt{3}R}{r\mathrm{sin}\theta}.
\end{aligned}
\end{equation}


\subsection{Numerical simulation}
In the real numerical simulations, to obtain a single vortex in each component, additional potential field constraints are required,
\begin{equation}
\begin{aligned}
     &i\hbar\partial_t\Psi_i=
     \big( -\frac{\hbar^2\nabla^2}{2m} +V_{trap}-\mu\\&+g|\Psi_i|^2+\eta(|\Psi_j|^2+|\Psi_k|^2)+\epsilon|\Psi_j|^2|\Psi_k|^2 \big)\Psi_i.
\end{aligned}
\end{equation}
We consider the general power-law trapping potential $V_{trap}({\bf r})=\frac{1}{2}mR^2_0(\frac{|{\bf r}|}{R_0})^\gamma$ . Then 
this potential field can be regarded as a circular infinite-depth potential well with $\gamma=100$, $R_0=m=1$. 
We employ Fourier methods to create two-dimensional periodic boundary conditions, with the number of grid points taken as $1000 \times 1000 (nx\times ny)$ and the boundary size taken as $30\times30 (Rx\times Ry)$. 
After placing the vortices using the aforementioned methods, the vortices and potential fields are first fully formed through imaginary time evolution. Subsequently, the fourth-order Runge-Kutta method is used for dynamic evolution.

{\it Acknowledgements.}
We would like to thank X. Wang, K. Y. Lu and Q. Y. Zhang for discussions.
This work was supported by the National Natural Science Foundation of China (under Grant No. 12275233),  JSPS KAKENHI Grant Number JP23K03305 and JP22H05139, and in part by JSPS KAKENHI (No. JP22H01221) and the WPI program ``Sustainability with Knotted Chiral Meta Matter (SKCM$^2$)'' at Hiroshima University.


%

\end{document}